\documentclass[a4paper,11pt]{article}

\usepackage{amsfonts,amssymb}
\usepackage{theorem}
\usepackage{epsf}

\def\G{\Gamma}

\begin{document}

%\rightline{Bern ... 20.6.2006}
%\rightline{revised  Milan ... 10.1.2007}

\medskip

\rightline{IFUM-884-FT}

\vskip 1 truecm
\Large
\bf
\centerline{Path-integral over non-linearly realized groups}
\centerline{and Hierarchy solutions}

\large
\rm
\vskip 1.3 truecm
\centerline{D.~Bettinelli\footnote{e-mail: 
{\tt daniele.bettinelli@mi.infn.it}}, 
R.~Ferrari\footnote{e-mail: {\tt ruggero.ferrari@mi.infn.it}}, 
A.~Quadri\footnote{e-mail: {\tt andrea.quadri@mi.infn.it}}}

\normalsize
\medskip
\begin{center}
Dip. di Fisica, Universit\`a degli Studi di Milano\\
and INFN, Sez. di Milano\\
via Celoria 16, I-20133 Milano, Italy
\end{center}

\vskip 0.8 truecm
\bf

\centerline{Abstract}

\normalsize
\rm

\vskip 0.5 truecm
\begin{quotation}
The technical problem of deriving 
the full Green functions of the
elementary pion fields of the nonlinear
sigma model in terms of ancestor
amplitudes involving only the flat connection
and the nonlinear sigma model constraint
is a very complex task.
In this paper we solve this problem by integrating, order by order
in the perturbative loop expansion,
the local functional equation derived from
the invariance of the SU(2) Haar measure under
local left multiplication. 
This yields  the
perturbative definition of the path-integral over
the non-linearly realized SU(2) group.
\end{quotation}
\newpage

\normalsize

\section{Introduction}

\setcounter{footnote}{0}

The perturbative quantization of the nonlinear
sigma model in $D=4$ requires a strategy for the
definition of the path-integral over the Haar measure
of non-linearly realized groups.

It has been recently pointed out \cite{Ferrari:2005ii}-
\cite{Ferrari:2005fc} that such a definition
can be implemented through the local functional equation
which expresses the invariance of the Haar measure 
under local left group multiplication.
The subtraction procedure is required to be symmetric,
thus preserving the validity of the local functional
equation to all orders in the loop expansion~\cite{Ferrari:2005fc}.
%In \cite{Ferrari:2005fc} it has been shown that the prescription of subtracting the poles after proper normalization
%of the amplitudes wtih a single mass scale $v_D$ 
%(the classical v.e.v. of the order parameter in $D$-dimensions)
%is indeed symmetric.

%In the loop expansion

The local functional equation fixes the Green functions
of the quantized pion fields parameterizing the SU(2)
group element (over which the path-integral is performed)
in terms of those of the SU(2) flat connection and the order
parameter (ancestor composite operators). 
This goes under the name of hierarchy principle
\cite{Ferrari:2005ii}.
Moreover there is only a finite number of divergent 
 ancestor amplitudes at every loop order
 (weak power-counting theorem \cite{Ferrari:2005va},\cite{Ferrari:2005fc}).
 
%The procedure is compatible with the
%ocality properties encoded in the weak power-counting
%theorem \cite{Ferrari:2005va},\cite{Ferrari:2005fc}.

The local solutions of the linearized  functional equation
(relevant for the classification of the
allowed finite renormalizations order by order
in the loop expansion)
were obtained in \cite{Ferrari:2005va}. 
In the one-loop approximation these results
	have been shown \cite{Ferrari:2005va}
to reproduce those of Ref.~\cite{Gasser:1983yg}.  

In this paper we show how to explicitly solve the local
functional equation by reconstructing the full Green functions
of the quantized fields once the relevant ancestor
amplitudes are known, to every order in the loop expansion.

In the one-loop approximation (linearized functional equation)
this is achieved by group-theoretical methods
allowing to introduce a suitable set of invariant
variables in one-to-one correspondence
with the external sources  
$J_{a\mu}$ (coupled  in the classical action
to the flat connection) and $K_0$
(coupled to the order parameter).
These invariant variables give rise to the whole dependence
of the one-loop vertex functional on the quantized
fields. 

As a special case one can apply this algorithm to the 
space of local functionals. We then show that 
the results of Ref.~\cite{Ferrari:2005va} are recovered.

At higher orders one has to solve an inhomogeneous
equation. For that purpose we make use of
algebraic BRST techniques originally
developed in the context of gauge theories
\cite{Ferrari:1998jy}-\cite{Quadri:2005pv}
 in order to invert the linearized operator in the relevant sector at ghost number one.

The main result is that starting from two loops on
the dependence of the vertex functional on the
quantized fields $\phi_a$ is two-fold: the $n$-th loop
ancestor amplitudes induce the dependence on the $\phi$'s
through the invariant variables solution
of the linearized functional equation
(implicit dependence). The lower-order contributions
(giving rise to the inhomogeneous term as a consequence
of the bilinearity of the functional equation) account
for the explicit dependence of
the $n$-th order vertex functional on the quantized fields.

We stress that in this approach
the functional equation is recursively solved order by order
in the loop expansion.
This allows to obtain the full dependence of
the vertex functional on the quantized fields
(which is uniquely determined once the ancestor amplitudes
are known) to all loop orders.

This algorithm can be applied to many problems arising
in the quantization of 
nonrenormalizable theories based on the hierarchy principle.
We just mention two of them here. 
The technique discussed in this paper can be applied
to higher loops Chiral Perturbation theory 
\cite{Bijnens:1999hw} in order
to determine the full dependence of the vertex
functional on the pion fields (including those 
terms which are on-shell vanishing).

Moreover this method is expected to provide a very useful tool
in the program of  the consistent quantization
of the Stueckelberg model \cite{Stueckelberg}-\cite{Ferrari:2004pd} for massive
non-abelian gauge bosons. 

\medskip
The paper is organized as follows.
In Sect.~\ref{sec.1} we briefly review the 
subtraction procedure based on the 
hierarchy principle in the flat connection formalism.
In Sect.~\ref{sec.2} we solve the local functional equation
in the one-loop approximation in full generality.
We do not impose  any locality restrictions on the space of the
solutions.
In  Sect.~\ref{sec:oneloop} we discuss some one-loop
examples. We show that by applying the
algorithm of Sect.~\ref{sec.2} to the space of local
functionals the results of Ref.~\cite{Ferrari:2005va} are recovered. 
We also solve explicitly the hierarchy for the four-point
pion amplitudes (one loop).
In Sect.~\ref{sec.3} the technique for the
determination of the higher order solution is developed.
In Sect.~\ref{sec:2loop} we apply this
technique on some examples at the two loop level.
In particular we obtain the solution of the hierarchy
for the four point pion functions at two loops.
In Sect.~\ref{sec:fr} we comment on the possible
finite renormalizations which are allowed
from a mathematical point of view by the weak
power-counting, order by order in the loop expansion,
and we show that they can be interpreted
as a redefinition of the external sources
$J_{a\mu}$ and $K_0$ by finite quantum
corrections.
Conclusions are finally given in Sect.~\ref{sec.5}.

%%%%%%%%%%%%%%%%%%%%%%%%%%%%%%%

\section{The flat connection formalism}\label{sec.1}

In the flat connection formalism \cite{Ferrari:2005ii}
the pion fields are embedded into the SU(2) flat connection
\begin{eqnarray}
&& F_\mu = i \Omega \partial_\mu \Omega^\dagger = \frac{1}{2} F_{a\mu} \tau_a \, .
\label{eq.flat}
\end{eqnarray}
In the above equation $\tau_a$ are the Pauli matrices
and $\Omega$ denotes the SU(2) group element. $\Omega$
is parameterized in terms of the pion fields $\phi_a$ 
as follows:
\begin{eqnarray}
&& \Omega = \frac{1}{v_D} (\phi_0 + i \tau_a \phi_a) \, , ~~~
\Omega^\dagger \Omega =1 \, , ~~~ {\rm det} ~ \Omega = 1\, ,
\nonumber \\
&&
\phi_0^2 + \phi_a^2 = v_D^2 \, .
\label{sec.1:1}
\end{eqnarray}
$v_D$ is the $D$-dimensional mass scale 
$$v_D = v^{D/2-1}$$
and $v$ has  mass dimension one. 

The $D$-dimensional action of the nonlinear sigma model
is written  in the presence of an external
vector source 
$J_{a\mu}$ \footnote{In this paper we denote by $J_{a\mu}$
the background connection. 
The classical action of Ref.~\cite{Ferrari:2005ii} differs
by a term $\frac{v_D^2}{8} J_{a\mu}^2$ w.r.t. 
the action in eq.(\ref{sec.1:2}). 
The source coupled to the flat
connection is given by $- \frac{v_D^2}{4} J_{a\mu}$.
Moreover we set the gauge coupling constant $g$
to $1$.}
and of a scalar source $K_0$ coupled to the
solution of the nonlinear sigma model constraint $\phi_0$:
\begin{eqnarray}
\G^{(0)} = \int d^Dx \, 
\Big ( \frac{v_D^2}{8} (F_{a\mu} - J_{a\mu})^2
      + K_0 \phi_0 \Big ) \, .
\label{sec.1:2}
\end{eqnarray}

The invariance of the Haar measure in the path-integral
under the local gauge transformations 
\begin{eqnarray}
&& \Omega ' = U \Omega \, , \nonumber \\
&& F'_\mu = U F_\mu U^\dagger + i U \partial_\mu U^\dagger \, , 
\label{sec.1:3}
\end{eqnarray}
where $U$ is an element of SU(2) allows to derive the following
local functional equation for the 1-PI vertex functional $\G$
\cite{Ferrari:2005ii}
\begin{eqnarray}
\Big ( - \partial_\mu \frac{\delta \G}{\delta J_{a \mu}}
       + \epsilon_{abc} J_{c\mu} \frac{\delta \G}{\delta J_{b \mu}} 
+ \frac{1}{2} K_0 \phi_a +  
\frac{1}{2} \frac{\delta \G}{\delta K_0} \frac{\delta \G}{\delta \phi_a}
+ \frac{1}{2} \epsilon_{abc} \phi_c \frac{\delta \G}{\delta \phi_b} 
%+ 2 D [ \frac{\delta \G}{\delta J} ]_{ab}^\mu J_{b\mu} 
\Big ) (x) = 0 \, .
\label{sec.1:4}
\end{eqnarray}
Moreover one requires that the vacuum expectation
value of the order parameter is fixed by the 
condition
\begin{eqnarray}
\left . \frac{\delta \G}{\delta K_0(x)} \right |_{\vec{\phi} = K_0 = J_{a\mu} = 0} = v_D \, .
\label{sec.norm.cond}
\end{eqnarray}

%
%This local functional equation has to be fulfilled
%by the subtraction procedure.

A weak power-counting theorem 
\cite{Ferrari:2005va} exists for the loop-wise perturbative 
expansion. Accordingly at any given loop order the number of
divergent ancestor amplitudes (i.e. those only involving
the insertion of the ancestor composite operators) is finite.
On the contrary,  already
at one loop level there is an infinite number of divergent 1-PI
amplitudes involving the $\phi_a$ fields (descendant amplitudes).
The latter can be fixed in terms of the ancestor ones by recursively
differentiating the local functional equation (\ref{sec.1:4}).

\section{One-loop solution}\label{sec.2}

In the one-loop approximation eq.(\ref{sec.1:4})
becomes
\begin{eqnarray}
\!\!\!\!\!\!
 {\cal S}_a (\G^{(1)}) & = & 
\Big ( - \partial_\mu \frac{\delta \G^{(1)}}{\delta J_{a \mu}}
+ \epsilon_{abc} J_{c\mu} \frac{\delta \G^{(1)}}{\delta J_{b \mu}}
+ \frac{1}{2} \frac{\delta \G^{(0)}}{\delta K_0} \frac{\delta \G^{(1)}}{\delta \phi_a}
+ \frac{1}{2} \frac{\delta \G^{(1)}}{\delta K_0} \frac{\delta \G^{(0)}}{\delta \phi_a}
\nonumber \\
& & 
+ \frac{1}{2} \epsilon_{abc} \phi_c \frac{\delta \G^{(1)}}{\delta \phi_b} 
\Big ) (x) = 0 \, .
\label{sec.1:6}
\end{eqnarray}
In order to solve the above equation we construct
invariant variables in one-to-one correspondence with 
the external sources. 
For that purpose we remark that the combination
\begin{eqnarray}
\overline{K}_0 = 
\frac{v_D^2 K_0}{\phi_0} - \phi_a \frac{\delta S_0}{\delta \phi_a}
\label{sec.1:9}
\end{eqnarray}
with
\begin{eqnarray}
S_0 = \frac{v_D^2}{8} \int d^Dx \, \Big ( F_{a\mu} - J_{a\mu} \Big )^2 
\label{sec.1:10}
\end{eqnarray}
is an invariant \cite{Ferrari:2005va}.
Moreover the transformation $K_0 \rightarrow \overline{K}_0$
is invertible.

On the other hand eq.(\ref{sec.1:6}) implies that
$J_{a\mu}$ transforms as a background connection.

The transformation properties of $\phi_a$
implement the 
non-linearly realized
SU(2) local transformation in eq.(\ref{sec.1:3}).
Hence $F_{a\mu}$ transforms as a gauge connection and therefore
the combination 
\begin{eqnarray}
I_\mu = I_{a\mu} \frac{\tau_a}{2} = F_{\mu} -  J_{\mu} 
\label{sec.1:8}
\end{eqnarray}
transforms in the adjoint representation (being
the difference of two connections):
\begin{eqnarray}
I'_\mu = U I_\mu U^\dagger \, .
\label{sec.1:8t}
\end{eqnarray}
As a consequence the conjugate of $I_\mu$ w.r.t. $\Omega$ 
\begin{eqnarray}
j_\mu = j_{a \mu}  \frac{\tau_a}{2} = \Omega^\dagger I_\mu \Omega 
\label{sec.1:n1}
\end{eqnarray} 
is 
an invariant under the transformations in
eqs.(\ref{sec.1:3}) and (\ref{sec.1:8t}).

By direct computation one finds that $j_{a \mu}$ in eq.(\ref{sec.1:n1}) is given by
\begin{eqnarray}
v_D^2 ~ j_{a\mu} & = & v_D^2 I_{a\mu} - 2 \phi_b^2 I_{a\mu} + 
                 2 \phi_b I_{b\mu} \phi_a 
                 + 2 \phi_0 \epsilon_{abc} \phi_b I_{c\mu} \nonumber \\
                 & \equiv & v_D^2 ~ R_{ba} I_{b\mu} 
                 \, .
\label{sec.1:n2}
\end{eqnarray}
%
%{\bf\Large Usando la notazione
%%
%%
%\begin{eqnarray}
% R_{ab} I_{b\mu} \equiv j_{a\mu}
%\label{sec.1:n2.1}
%\end{eqnarray}
%%
%ottengo
%%
%%
%\begin{eqnarray}
%v^2_{D}~~ R_{ab} = (v^2_{D}-2 \vec \phi^2)\delta_{ab}+
%2\phi_a\phi_b - 2 \phi_0\phi_c \epsilon_{abc}
%\label{sec.1:n2.2}
%\end{eqnarray}
%%
%che va d'accordo con la (\ref{sec.1:n2}).
%}

The matrix $R_{ba}$ in the above equation
is an element of the adjoint representation of the SU(2) group.
Hence the transformation $J_{a\mu} \rightarrow j_{a\mu}$ is invertible.

%{\bf\Large Sono scarso in teoria dei gruppi, ma mi sembra che questa
%affermazione non sia corretta:
%%
%\begin{eqnarray}
%&&
%J=0 \quad \frac{1}{3}\delta_{ab}R_{cc}=\delta_{ab}(v^2_{D}-2 \vec \phi^2+
%\frac{2}{3} \phi^2)
%\nonumber\\ &&
%J= 2 \quad \frac{1}{2}( R_{ab}+ R_{ab})- \frac{1}{3}\delta_{ab}R_{cc}
%\nonumber\\ &&
%J=1 \quad 2 \phi_0\phi_c \epsilon_{abc}
%\label{sec.1:n2.3}
%\end{eqnarray}
%
%Forse si vuol dire che l'indice $a$ non trasfomato implica che $b$ \'e
%un indice vettoriale?

%E' importante sapere come si trasforma $R_{ab}$ altrimenti non \'e chiaro
%cosa vuol dire la (\ref{ex.1}).
%}

The linearized functional equation (\ref{sec.1:6})
has a very simple form in the variables $\{ \phi_a, \overline{K}_0, j_{a\mu} \}$.
In fact, by taking into account the invariance of $\overline{K}_0$ and $j_{a\mu}$
under ${\cal S}_a$, 
eq.(\ref{sec.1:6}) reduces to
\begin{eqnarray}
%&& \Big ( \frac{1}{2} \phi_0 z\delta_{ab} + \frac{1}{2} \epsilon_{abc} \phi_c \Big ) 
%\frac{\delta \G^{(1)}[\overline K_0, j_{a\mu}, \phi_a]}{\delta \phi_b(x)} \nonumber \\
%&& ~~~~ \equiv
\Theta_{ab}\frac{\delta\G^{(1)}[\phi_a, \overline K_0, j_{a\mu}]}{\delta \phi_b} = 0 \, , 
\label{sec.1:n3}
\end{eqnarray}
where the matrix $\Theta_{ab}$ gives the variation of $\phi_b$:
\begin{eqnarray}
\Theta_{ab} =  \frac{1}{2} \phi_0 \delta_{ab} + \frac{1}{2} \epsilon_{abc} \phi_c \, .
\label{sec.1:n4}
\end{eqnarray}
$\Theta_{ab}$ is invertible as a consequence of the nonlinear constraint in the
second line of eq.(\ref{sec.1:1}) and thus eq.(\ref{sec.1:n3}) is equivalent to
\begin{eqnarray}
\frac{\delta\G^{(1)}[\phi_a, \overline K_0, j_{a\mu}]}{\delta \phi_b} = 0 \, .
\label{sec.1:n5}
\end{eqnarray}
That means that the only dependence of the symmetric vertex functional
$\G^{(1)}$ on the pion fields is through the variables $\overline{K}_0$ and $j_{a\mu}$.

This in turn allows to integrate the linearized functional equation (\ref{sec.1:6}).
For that purpose
one has to replace in the ancestor amplitudes 1-PI functional the source $K_0$ with
$\frac{1}{v_D} \overline{K}_0$ and $J_{a\mu}$ with $-j_{a\mu}$. The normalization
of $\overline{K}_0$ and $j_{a\mu}$ is fixed by the 
boundary conditions
\begin{eqnarray}&&
\overline K_0 |_{\vec\phi=0} = v_D K_0
\nonumber\\&&
- j_{a\mu}|_{\vec\phi=0} = J_{a\mu} \, .
\label{sec.1:n5.1}
\end{eqnarray}

By eq.(\ref{sec.1:n5}) this algorithm gives rise to the full dependence on
the pion fields at the one loop level.
Thus we can state the following Proposition:
\medskip

{\bf Proposition 1}. Given the 
ancestor amplitudes 1-PI functional ${\cal A}^{(1)}[K_0,J_{a\mu}]$
the solution of the linearized local functional equation (\ref{sec.1:6}) is obtained through the replacement rule
\begin{eqnarray}
\G^{(1)}[\phi_a,K_0,J_{a\mu}] = 
\left .
{\cal A}^{(1)}[K_0,
J_{a\mu}] \right |_{K_0 \rightarrow
\frac{1}{v_D} \overline{K}_0, J_{a\mu} \rightarrow -j_{a \mu}}
\, 
\label{sec4.sol.NL}
\end{eqnarray}
where in the R.H.S. of the above equation
$\overline{K}_0$ is given by eq.(\ref{sec.1:9}) and $j_{a \mu}$ by eq.(\ref{sec.1:n2}).

In view of this result we say that $\G^{(1)}$ depends on the $\phi$'s only
implicitly (i.e. through $\overline{K}_0$ and $j_{a\mu}$).
This terminology will prove convenient when
studying the dependence of the vertex functional
on the $\phi$'s at higher orders.

\medskip
We stress that no restriction to the space of local
functionals is used in the above derivation.
Eq.(\ref{sec4.sol.NL}) thus provides the full set
of Green functions involving at least one pion
in terms of the ancestor amplitudes. This solves
the hierarchy at the one loop level.

\section{One-loop examples}\label{sec:oneloop}

When restricted to the local (in the
sense of formal power series) functionals, 
the prescription
in eq.(\ref{sec4.sol.NL}) gives back
the results of \cite{Ferrari:2005va}.
This follows from the uniqueness of the hierarchy solution once
the ancestor amplitudes are fixed.

As an example we derive the local invariants ${\cal I}_1, \dots,{\cal I}_7$
parameterizing the one-loop divergences of the nonlinear sigma model in $D=4$
(see Appendix~\ref{app:B})
by performing the substitution 
$K_0 \rightarrow \frac{1}{v_D} \overline{K}_0$ , $J_{a\mu} \rightarrow -j_{a\mu}$ 
in the relevant ancestor monomials
\begin{eqnarray}
& 
\int %d^Dx 
      \, \partial_\mu J_{a\nu} \partial^\mu J^\nu_a\, , ~~~
\int %d^Dx 
      \, \partial J_a \partial J_a \, , ~~~
\int %d^Dx 
      \, \epsilon_{abc} \partial_\mu J_{a \nu} J^\mu_b J^\nu_c \, , &
\nonumber \\
& \int %d^Dx 
      \, K_0^2 \, , ~~~
  \int %d^Dx 
      \, K_0 J^2 \, ,  ~~~
  \int %d^Dx 
      \, (J^2)^2 \, , ~~~
  \int %d^Dx 
      \, J_{a\mu} J^\mu_b J_{a \nu} J^\nu_b \, .
&
\label{ex.10}
\end{eqnarray}
The monomials in the second line of the above equation do not contain derivatives. By using the SU(2)
constraint
\begin{eqnarray}
R_{ba} R_{ca} = \delta_{bc}
\label{ex.12}
\end{eqnarray}
we get 
\begin{eqnarray}
&& j_{a\mu}^2 %= \Omega_{ba} I_{b\mu} \Omega_{ca} I_c^\mu 
 = I_{a\mu}^2 \, , ~~~~  %\nonumber \\
%&& 
j_{a\mu} j^\mu_b j_{a \nu} j^\nu_b %= \Omega_{pa} \Omega_{qa} \Omega_{rb} \Omega_{sb} I_{p\mu} I_{r}^\mu I_{q\nu} I^\nu_s 
= I_{a\mu} I_{b}^\mu I_{a\nu} I^\nu_b \, .
\label{ex.12.1}
\end{eqnarray}
Therefore
\begin{eqnarray}
&& \int d^Dx \, K_0^2  \rightarrow \frac{1}{v_D^2} \int d^D x \, \overline{K}_0^2 = \frac{1}{v_D^2} {\cal I}_4 \, , 
\nonumber \\
&& \int d^Dx \, K_0 J^2 \rightarrow 
\frac{1}{v_D} \int d^Dx \, \overline{K}_0 j^2 = \frac{1}{v_D} {\cal I}_5 \, , 
\nonumber \\
&&  \int d^Dx 
      \, (J^2)^2 \,  \rightarrow \int d^Dx \, (j^2)^2 = {\cal I}_6 \, ,
\nonumber \\
&&  \int d^Dx 
      \, J_{a\mu} J^\mu_b J_{b \nu} J^\nu_b 
      \rightarrow \int d^Dx \, j_{a\mu} j^\mu_b j_{b \nu} j^\nu_b = {\cal I}_7 \, . 
\label{ex.13}
\end{eqnarray}

In order to establish the matching for the ancestor monomials involving derivatives
in the first line of eq.(\ref{ex.10}), we notice that 
the flat connection $F_{a\mu}$ can be computed in terms of $R_{ba}$ as well
(since $R_{ba}$ belongs to the adjoint representation of the SU(2) group). In fact one finds 
\begin{eqnarray}
i R_{bc} \partial_\mu R^\dagger_{ca} = i R_{bc} \partial_\mu R_{ac} =
(T_c)_{ba} F_{c\mu} 
\label{ex.1}
\end{eqnarray}
where $(T_c)_{ba} = i \epsilon_{cab}$ are the generators of the adjoint representation
satisfying the commutation relation
\begin{eqnarray}
[T_a, T_b] = i \epsilon_{abc} T_c \, .
\label{ex.2}
\end{eqnarray}
%
%%%%%%%%%%%%%%%%%%%%%%%%%%%%%%%%%%%%%%%%%%%%%%%%%%%%%%%%%%%%%%%%%%%
Eq.(\ref{ex.1}) can be checked  as follows.
We set
\begin{eqnarray}&&
R_a \equiv \Omega^\dagger \tau_a\Omega = \tau_bR_{ab}
\nonumber \\&&
R_{ab}= \frac{1}{2} T_r\left( \tau_b \Omega^\dagger \tau_a\Omega
\right) \, .
\label{or.1}
\end{eqnarray}
By using the following identities
\begin{eqnarray}&&
T_r\left( \tau_a F_\mu \tau_b 
\right)
= T_r\left(  \Omega^\dagger\tau_a F_\mu \tau_b \Omega
\right)
\nonumber \\&&
=i T_r\left(  \Omega^\dagger\tau_a\Omega \partial_\mu \Omega^\dagger\tau_b \Omega
\right)
\nonumber \\&&
=i T_r\left(  \Omega^\dagger\tau_a\Omega \partial_\mu \left[\Omega^\dagger\tau_b 
\Omega\right]\right)
-i T_r\left(  \Omega^\dagger\tau_a\Omega  \Omega^\dagger\tau_b \partial_\mu\Omega
\right)
\nonumber \\&&
=i T_r\left(  \Omega^\dagger\tau_a\Omega \partial_\mu \left[\Omega^\dagger\tau_b 
\Omega\right]\right)
+i T_r\left(  \tau_a\Omega  \Omega^\dagger\tau_b \Omega\partial_\mu\Omega^\dagger
\right)
\label{or.2}
\end{eqnarray}
we find
\begin{eqnarray}
T_r\left( \tau_a F_\mu \tau_b 
\right)-T_r\left( \tau_b F_\mu \tau_a 
\right)
=i T_r\left(  \Omega^\dagger\tau_a\Omega \partial_\mu \left[\Omega^\dagger\tau_b 
\Omega\right]\right)
\label{or.3}
\end{eqnarray}
which gives directly eq.(\ref{ex.1}):
\begin{eqnarray}
-i\epsilon_{abc} F_{c\mu}
=i  R_{ac}\partial_\mu R_{bc} 
= - i R_{bc} \partial_\mu R_{ac} \, .
\label{or.4}
\end{eqnarray}
%
%%%%%%%%%%%%%%%%%%%%%%%%%%%%%%%%%%%%%%%%%%%%%%%%%%%%%%%%%%%%%%%%%%%5555

By repeated application of eq.(\ref{ex.12}) and eq.(\ref{ex.1}) we then get
\begin{eqnarray}
&& \int d^Dx \, \partial_\mu J_{a\nu} \partial^\mu J^\nu_a \rightarrow 
   \int d^Dx \, \partial_\mu j_{a\nu} \partial^\mu j^\nu_a %\nonumber \\
%&& ~~~~~~~~~~~~ 
= \int d^Dx \, \partial_\mu \Big ( R_{ba} I_{\nu b} \Big ) \partial^\mu \Big ( R_{ca} I^\nu_c \Big ) \nonumber \\
&& ~~~~~~~~~~~~ = \int d^Dx \, (D_\mu[F] I_\nu)_a (D^\mu[F] I^\nu)_a  = {\cal I}_1 \, ,
\label{ex.15}
\end{eqnarray}
where $D_\mu[F]$ is the covariant derivative w.r.t. $F_{a\mu}$:
\begin{eqnarray}
(D_\mu[F] I_{\nu})_a = \partial_\mu I_{a\nu} + \epsilon_{abc} F_{b \mu} I_{c\nu} \, .
\label{ex.16}
\end{eqnarray}
In a similar way we get
\begin{eqnarray}
&& \!\!\!\!\!\!\!\!\!\!\!\!\!\!\!\!
\int d^Dx \, \partial J_a \partial J_a \rightarrow \int d^Dx \, \partial j_a \partial j_a %\nonumber \\
%&& ~~~~ 
%= \int d^Dx \, \partial_\mu \Big ( \Omega_{ba} I_{\nu b} \Big ) \partial^\mu \Big ( \Omega_{ca} I^\nu_c \Big ) %\nonumber \\
= \int d^Dx \, (D_\mu[F] I^\mu)_a (D_\nu[F] I^\nu)_a = {\cal I}_2 \, . \nonumber \\
\label{ex.17}
\end{eqnarray}
Moreover
\begin{eqnarray}
&& 
\!\!\!\!\!\!\!\!\!\!\!\!\!\!\!\!\!\!\!\!\!\!\!\!\!\!\!\!\!
\int d^Dx \, \epsilon_{abc} \partial_\mu J_{a \nu} J^\mu_b J^\nu_c \rightarrow 
- \int d^Dx \, \epsilon_{abc} \partial_\mu j_{a \nu} j^\mu_b j^\nu_c \nonumber \\
&& \!\!\!\!\!\!\!\!\!\! = - \int d^Dx \, \epsilon_{abc} \Big ( \partial_\mu R_{qa} I_{q \nu}
                          R_{pb} I^\mu_p R_{rc} I^\nu_r %\nonumber \\
%&& ~~~~~~~~~  
%- \int d^Dx \, \epsilon_{abc} 
+ R_{qa} \partial_\mu I_{q\nu} R_{pb} I^\mu_p R_{rc} I^\nu_r \Big ) \, .
\label{ex.18}
\end{eqnarray}
By noticing that 
\begin{eqnarray}
\epsilon_{abc} R_{qa} R_{pb} R_{rc} = \epsilon_{qpr}
\label{ex.19}
\end{eqnarray}
and by using eqs.(\ref{ex.12}) and (\ref{ex.1}) into eq.(\ref{ex.18}) we finally get
\begin{eqnarray}
- \int d^Dx \, \epsilon_{abc} \partial_\mu j_{a \nu} j^\mu_b j^\nu_c = 
- \int d^Dx \, \epsilon_{abc} (D_\mu[F] I_\nu)_a I^\mu_b I^\nu_c = - {\cal I}_3 \, .
\label{ex.20}
\end{eqnarray}

\medskip
As we have mentioned several times, the algorithm for solving the hierarchy based on Proposition 1
can be applied in order to derive the full Green functions involving at least one pion field in terms
of the ancestor amplitudes. 
As an example, we obtain here the full one-loop four point pion amplitude in terms of the relevant
ancestor amplitudes $\G^{(1)}_{JJ}, \G^{(1)}_{JJJ}, \G^{(1)}_{JJJJ}, \G^{(1)}_{K_0 K_0}$ and 
$\G^{(1)}_{K_0 JJ}$.
For that purpose one has to perform the substitution $J_{a\mu} \rightarrow -j_{a\mu}$ and 
$K_0 \rightarrow \frac{1}{v_D} \overline{K}_0$ in the relevant part of the ancestor functional
\begin{eqnarray}
&& 
\!\!\!\!\!\!\!\!\!
{\cal A}^{(1)}[K_0, J_{a\mu} ]  =  \frac{1}{2} \int %d^Dx d^Dy 
\,  \G^{(1)}_{J_{a\mu}(x) J_{b\nu}(y)} J_{a\mu}(x) J_{b\nu}(y) 
+ \nonumber \\
&& + \frac{1}{3!} \int %d^Dx d^Dy d^Dz 
\,  \G^{(1)}_{J_{a\mu}(x) J_{b\nu}(y) J_{c\rho}(z)} J_{a\mu}(x) J_{b\nu}(y) J_{c\rho}(z)
\nonumber \\ 
&& + \frac{1}{4!} \int %d^Dx d^Dy d^Dz d^Dw 
\,  \G^{(1)}_{J_{a\mu}(x) J_{b\nu}(y) J_{c\rho}(z) J_{d\sigma}(w)} J_{a\mu}(x) J_{b\nu}(y) J_{c\rho}(z) J_{c\sigma}(w) \nonumber \\
&& + \frac{1}{2} \int %d^Dx d^Dy d^Dz 
\,  \G^{(1)}_{J_{a\mu}(x) J_{b\nu}(y) K_0(z)} J_{a\mu}(x) J_{b\nu}(y) K_0(z) \nonumber \\
&& + \frac{1}{2} \int %d^Dx d^Dy 
\,  \G^{(1)}_{K_0(x) K_0(y)} K_0(x) K_0(y) + \dots 
\label{ex.20.1}
\end{eqnarray}
by keeping
all terms contributing up to four pion fields. This amounts to truncate the expansion of $\overline{K}_0$
up to two $\phi$'s and the expansion of $j_{a\mu}$ up to three $\phi$'s:
\begin{eqnarray}
&& 
\!\!\!\!\!\!\!\!\!\!\!\!\!\!\!\!\!\!\!\!\!\!
\overline{K}_0 = \frac{1}{v_D} \phi_a \square \phi_a + \dots \, , \nonumber \\
&& 
\!\!\!\!\!\!\!\!\!\!\!\!\!\!\!\!\!\!\!\!\!\!
j_{a\mu} = \frac{2}{v_D} \partial_\mu \phi_a - \frac{2}{v_D^2} \epsilon_{abc} \partial_\mu \phi_b \phi_c
+ \frac{1}{v_D^3} \Big ( - \phi_b^2 \partial_\mu \phi_a + 2 \phi_b \partial_\mu \phi_b \phi_a \Big ) + \dots 
\label{ex.21}
\end{eqnarray}
Then one gets
\begin{eqnarray}
&& 
\!\!\!\!\!\!\!\!\!\!\!\!\!\!\!\!\!\!\!\!\!\!\!\!\!\!\!\!\!\!\!\!\!\!\!\!\!
\G^{(1)}[\phi\phi\phi\phi] = \frac{2}{v_D^4} \int %d^Dx d^Dy 
\, \G^{(1)}_{J_{a\mu}(x) J_{b\nu}(y)} 
\Big (  \partial_\mu \phi_a(x) (  - \phi_c^2(y) \partial_\nu \phi_b(y) + 2 \phi_c(y) \partial_\nu \phi_c(y) \phi_b(y)  )
\nonumber \\
&& ~~~~~~~~~~~~~~~~~~~~~~~ + \epsilon_{apq} \epsilon_{brs} \partial_\mu \phi_p(x) \phi_q(x) \partial_\nu \phi_r(y) \phi_s(y) \Big ) \nonumber \\
&& \!\!\!\!\!\! 
+ \frac{4}{v_D^4} \int %d^Dx d^Dy d^Dz 
\,  \G^{(1)}_{J_{a\mu}(x) J_{b\nu}(y) J_{c\rho}(z)}
\epsilon_{apq} \partial_\mu \phi_p(x) \phi_q(x) \partial_\nu \phi_b(y) \partial_\rho \phi_c(z) \nonumber \\
&& \!\!\!\!\!\!
+ \frac{2}{3 v_D^4}  \int %d^Dx d^Dy d^Dz d^Dw 
\,  \G^{(1)}_{J_{a\mu}(x) J_{b\nu}(y) J_{c\rho}(z) J_{d\sigma}(w)}
\partial_\mu \phi_a(x) \partial_\nu \phi_b(y) \partial_\rho \phi_c(z) \partial_\sigma \phi_d(w) \nonumber \\
&& \!\!\!\!\!\!
+ \frac{2}{v_D^3} \int \,  \G^{(1)}_{J_{a\mu}(x) J_{b\nu}(y) K_0(z)}\partial_\mu \phi_a(x) \partial_\nu \phi_b(y)
(\phi_c \square \phi_c)(z) \nonumber \\
&& \!\!\!\!\!\!
+ \frac{1}{2v_D^2} \int \, \G^{(1)}_{K_0(x) K_0(y)} (\phi_a \square \phi_a)(x) (\phi_b \square \phi_b)(y)
\, .
\label{ex.4pphi}
\end{eqnarray}

\section{Higher orders}\label{sec.3}

At higher orders the functional equation (\ref{sec.1:4}) 
yields an inhomogeneous
equation for $\G^{(n)}$, $n>1$:
\begin{eqnarray}
{\cal S}_a(\G^{(n)}) = - \frac{1}{2} \sum_{j=1}^{n-1}
\frac{\delta \G^{(j)}}{\delta K_0} \frac{\delta \G^{(n-j)}}{\delta \phi_a} \, .
\label{ho.0}
\end{eqnarray}

In order to recursively integrate 
eq.(\ref{ho.0}) order by order in the loop expansion
it is convenient to introduce a BRST formulation 
for the linearized functional
operator ${\cal S}_a$. For that purpose we define
the BRST differential~\cite{Ferrari:2005va}
$$ s = \int d^Dx \, \omega_a {\cal S}_a $$
where $\omega_a$ are classical anticommuting local parameters.
The variables $\overline{K}_0$ and $j_{a\mu}$ are $s$-invariant
while 
\begin{eqnarray}
s \phi_a = \frac{1}{2} \phi_0 \omega_a + \frac{1}{2} \epsilon_{abc} \phi_b \omega_c \equiv \Theta_{ab} \omega_b \, , ~~~~ s\omega_a = -\frac{1}{2} \epsilon_{abc} \omega_b \omega_c \, .
\label{ho.1}
\end{eqnarray}
The BRST transformation of $\omega_a$ is dictated by nilpotency.
$\omega_a$
have ghost number one, while all the remaining
variables have ghost number zero.
In view of the fact that there are no variables
with negative ghost number and that the vertex functional $\G$ has ghost
number zero, $\G$ cannot depend on $\omega_a$.

The introduction of a BRST differential allows
to make use of the technique of the Slavnov-Taylor (ST)
parameterization of the effective action \cite{Quadri:2003ui}-\cite{Quadri:2005pv}
(originally developed in order to restore the ST identities
for power-counting renormalizable gauge theories in the absence of a 
symmetric regularization) in order to solve the local functional equation
at orders $\geq 1$. 

For that purpose we remark that,
since the matrix $\Theta_{ab}$ in eq.(\ref{ho.1}) is invertible, we can 
perform a further
change of variables by setting
\begin{eqnarray}
\overline{\omega}_a = \Theta_{ab} \omega_b \, .
\label{ho.2}
\end{eqnarray}
The inverse matrix $\Theta^{-1}_{ca}$ is given by
\begin{eqnarray}
&& \Theta^{-1}_{ca} = \frac{2\phi_0}{v_D^2} \delta_{ca}
+ \frac{2}{v_D^2 \phi_0} \phi_c \phi_a 
- \frac{2}{v_D^2} \epsilon_{cpa} \phi_p \, .
\label{ho.3.1}
\end{eqnarray}
 The action of $s$ on the variables 
$\{ \overline{K}_0, j_{a\mu}, \phi_a, \overline{\omega}_a \}$
is finally given by
\begin{eqnarray}
&&
s  \overline{K}_0 = s j_{a \mu} = 0 \, , 
\nonumber \\
&&
s \phi_a = \overline{\omega}_a \, , ~~~~ s \overline{\omega}_a = 0 \, ,
\label{ho.3}
\end{eqnarray}
i.e. $s$ has been cohomologically trivialized: $(\phi_a, \overline{\omega}_a)$
form a BRST doublet \cite{Piguet:1995er}-\cite{Quadri:2002nh}, while  $\overline{K}_0$ and $j_{a\mu}$
are invariant.

We are now in a position to recursively solve the local functional equation
at higher orders in perturbation theory. 
By using the BRST differential $s$ eq.(\ref{ho.0}) reads
%The projection
%at order $n$ in the loop expansion yields the following
%equation for the $n$-th order vertex functional $\G^{(n)}$:
%
\begin{eqnarray}
s \G^{(n)} = \Delta^{(n)} \equiv 
- \frac{1}{2} \sum_{j=1}^{n-1} \int d^Dx \, \omega_a 
\frac{\delta \G^{(j)}}{\delta K_0(x)} \frac{\delta \G^{(n-j)}}{\delta \phi_a(x)}
\, ,
\label{sho.1}
\end{eqnarray}
where $\Delta^{(n)}$ depends only on known lower order terms.
Nilpotency of $s$ implies that $\Delta^{(n)}$ is invariant:
\begin{eqnarray}
s \Delta^{(n)} = 0 \, .
\label{ho.cc}
\end{eqnarray}

This consistency condition can be checked to hold as a consequence
of the fulfillment of the functional equation up to order $n-1$, as
shown in Appendix~\ref{appA}.

By using eq.(\ref{ho.3}) into eq.(\ref{sho.1}) we find
\begin{eqnarray}
\int d^Dx \, \overline{\omega}_a \frac{\delta \G^{(n)}}{\delta \phi_a} 
= \Delta^{(n)}[\overline{\omega}_a, \phi_a, \overline{K}_0, j_{a\mu}] \, .
\label{new.ho.2} 
\end{eqnarray}
We remark that $\Delta^{(n)}$ is linear in $\overline{\omega}_a$. By differentiating
eq.(\ref{new.ho.2}) and by setting $\overline{\omega}_a=0$ we get
\begin{eqnarray}
 \frac{\delta \G^{(n)}}{\delta \phi_a(x)} =  
\frac{\delta \Delta^{(n)}}{\delta \overline{\omega}_a(x)} 
\label{new.ho.3}
\end{eqnarray}
which fixes the explicit dependence of the symmetric
vertex functional $\G^{(n)}$ on $\phi_a(x)$ ($\G^{(n)}$ depends on $\phi$
also implicitly through the invariant variables $j_{a\mu}$ and $\overline{K}_0$).
By successive differentiation of eq.(\ref{new.ho.3})
we obtain
\begin{eqnarray}
\!\!\!\!\!\!\!\!\!\!\!\!\!
\G^{(n)}_{\phi_{a_1} \dots \phi_{a_m} \zeta_{b_1} \dots \zeta_{b_n}} 
& = & 
\Delta^{(n)}_{\overline{\omega}_{a_1} \phi_{a_2} \dots \phi_{a_m}
\zeta_{b_1} \dots \zeta_{b_n}} 
\nonumber \\
& = & \frac{1}{m!} \sum_{\sigma \in S_m} 
\Delta^{(n)}_{\overline{\omega}_{a_{\sigma(1)}} 
              \phi_{a_{\sigma(2)}} \dots 
              \phi_{a_{\sigma(m)}} 
              \zeta_{b_1} \dots \zeta_{b_n} }  
\label{new.ho.3.1}
\end{eqnarray}
where $\zeta$ is a collective notation 
standing for $j_{a\mu}$ and $\overline{K}_0$.

The equality in the second line of the above equation
is a consequence of the Bose statistics of the $\phi$'s.
We point out that eq.(\ref{new.ho.3.1}) imposes 
a consistency condition on $\Delta^{(n)}$, i.e.
\begin{eqnarray}
\!\!
\Delta^{(n)}_{\overline{\omega}_{a_1} \phi_{a_2} \dots \phi_{a_m}
              \zeta_{b_1} \dots \zeta_{b_n} }
= \frac{1}{m!} \sum_{\sigma \in S_m} 
\Delta^{(n)}_{\overline{\omega}_{a_{\sigma(1)}} 
              \phi_{a_{\sigma(2)}} \dots 
              \phi_{a_{\sigma(m)}}
              \zeta_{b_1} \dots \zeta_{b_n}}  \, . 
\label{new.ho.3.2}
\end{eqnarray}
This condition holds as a consequence of eq.(\ref{ho.cc}),
as is proven in Appendix~\ref{app:int}.

Eq.(\ref{new.ho.3}) shows that at order $n\geq 2$
the vertex functional exhibits a further dependence
on the $\phi$'s  (in addition to the implicit one
through the variables $\overline{K}_0$ and $j_{a\mu}$). 
We refer to it as the explicit dependence of $\G^{(n)}$
on $\phi_a$.
It is a remarkable fact that 
this latter dependence
%of $\G^{(n)}$
 on the pion fields comes from  amplitudes involving the pion field of lower order strictly.
In particular, they do not
affect the $n$-th loop ancestor amplitudes.

In order to recover the full $n$-th loop vertex
functional one also needs to take into account the 
implicit dependence on the pion fields through
$\overline{K}_0$ and $j_{a\mu}$. 
In fact we can state the following
\vskip 0.5 truecm
{\bf Proposition 2.} Given the functional ${\cal A}^{(n)}[K_0,J_{a\mu}]$
collecting the $n$-th order ancestor amplitudes, $n\geq2$, 
 the full $n$-th loop vertex functional is given by
\begin{eqnarray}
&& 
\G^{(n)}[\phi_a, K_0, J_{a\mu} ] = 
\left .
{\cal A}^{(n)}[K_0,
J_{a\mu}] \right |_{K_0 \rightarrow
\frac{1}{v_D} \overline{K}_0, J_{a\mu} \rightarrow -j_{a \mu}}
\nonumber \\
%{\cal A}^{(n)}[\frac{1}{v_D} \overline{K}_0, -j_{a\mu}]
&& ~~~~~~~~~~~~~~~~~~~~~~~~
+ \int d^Dx \, \int_0^1 dt \, \phi_a(x) \lambda_t \frac{\delta  \Delta^{(n)}}{\delta \overline{\omega}_a(x)} 
\label{new.ho.4}
\end{eqnarray}
where $\lambda_t$ acts as follows on a functional $X[\phi_a, \overline{K}_0, 
j_{a\mu}]$:
\begin{eqnarray}
\lambda_t X[\phi_a, \overline{K}_0, j_{a\mu}] = 
X[t \phi_a,  \overline{K}_0, j_{a\mu}] \, .
\label{new.ho.5}
\end{eqnarray}
The first term in the R.H.S. of eq.(\ref{new.ho.4}) 
accounts for the implicit dependence on $\phi_a$
through $\overline{K}_0$ and $j_{a\mu}$.
It is of the same form as in the one loop approximation
eq.(\ref{sec4.sol.NL}). 

The second term in the R.H.S. of eq.(\ref{new.ho.4}) 
is present only from two loops on. It arises as
a consequence of the bilinearity of the local
functional equation (\ref{sec.1:4}).
It gives rise 
to the explicit dependence of $\G^{(n)}$ on $\phi_a$
dictated by eq.(\ref{new.ho.3}).
This can be checked by taking derivatives w.r.t. $\phi_{a_1}, \dots, \phi_{a_m}$ of
eq.(\ref{new.ho.4}) and then setting $\phi=0$ (derivatives
w.r.t. $\zeta_{b_1}, \dots, \zeta_{b_n}$ do not play any role in the following argument). The only contribution
comes from the second term and yields
\begin{eqnarray}
&& 
\!\!\!\!\!\!\!
\frac{\delta^m}{\delta \phi_{a_1} \dots \delta \phi_{a_m}}
\int d^Dx \int_0^1 dt \, \phi_a(x) \lambda_t \frac{\delta \Delta^{(n)}}{\delta \overline \omega_a(x)} \nonumber \\
&& = 
\frac{\delta^m}{\delta \phi_{a_1} \dots \delta \phi_{a_m}} \int d^Dx d^Dy_1 \dots d^Dy_{m-1}\nonumber \\
&& ~~~~~~ 
\frac{1}{(m-1)!}  \int_0^1 dt~  t^{m-1} \Delta^{(n)}_{\overline \omega_a(x) \phi_{b_1}(y_1) \dots \phi_{b_{m-1}}(y_{m-1})} 
 \phi_a(x) \phi_{b_1}(y_1) \dots \phi_{b_{m-1}}(y_{m-1}) \nonumber \\
&& = \frac{1}{m!} \sum_{\sigma \in S_m} 
\Delta^{(n)}_{\overline{\omega}_{a_{\sigma(1)}} 
              \phi_{a_{\sigma(2)}} \dots 
              \phi_{a_{\sigma(m)}}} 
= \G^{(n)}_{\phi_{a_1} \dots \phi_{a_m}}              
\label{new.ho.bis}
\end{eqnarray}
where in the last line we have used  eq.(\ref{new.ho.3.1}).

Eq.(\ref{new.ho.4}) provides the full set of
$n$-th order Green functions in terms of $n$-th order
ancestor amplitudes and known lower order terms, thus
solving the hierarchy.

\section{Two-loop examples}\label{sec:2loop}

In this Section we apply the method developed
in Sect.~\ref{sec.3} at two loop order. 
The two-loop inhomogeneous term is 
\begin{eqnarray}
\Delta^{(2)} = - \int d^Dx \, \frac{1}{2} \omega_a(x) \frac{\delta \G^{(1)}}{\delta K_0(x)} \frac{\delta \G^{(1)}}{\delta \phi_a(x)} \, .
\label{EX.2l.1}
\end{eqnarray}
In order to apply eq.(\ref{new.ho.4}) we need to express
the R.H.S. in terms of the variables $\{ \overline{K}_0,
j_{a\mu}, \phi_a \}$. For that purpose we write
\begin{eqnarray}
\!\!\!\!\!\!\!\!\!\!\!\!\!\!\!
\Delta^{(2)} & = & - \int d^Dx \,
\frac{1}{2} \Theta^{-1}_{ab} \overline{\omega}_b \int d^Dy \, 
\frac{\delta \overline{K}_0(y)}{\delta K_0(x)}
\frac{\delta \G^{(1)}}{\delta \overline{K}_0(y)}  \nonumber \\
& & 
\!\!\!\!\!\!\!\!\!\!\!\!\!\!\!\!\!\!\!\!\!\!\!\!\!\!
\int d^Dz \, \Big ( 
\frac{\delta \overline{K}_0(z)}{\delta \phi_a(x)} 
\frac{\delta}{\delta \overline{K}_0(z)}
+
\frac{\delta j_{c\mu}(z)}{\delta \phi_a(x)} 
\frac{\delta}{\delta j_{c\mu}(z)} 
+
\delta^D(x-z) \frac{\delta}{\delta \phi_a(z)} \Big )
\G^{(1)} \, .
\label{EX.2l.2}
\end{eqnarray}
where the matrix $ \Theta^{-1}_{ab}$
is given in eq.(\ref{ho.3.1}).

Moreover
\begin{eqnarray}
\frac{\delta \overline{K}_0(y)}{\delta K_0(x)} = 
\frac{v_D^2}{\phi_0} \delta^D(y-x) 
\label{EX.2l.3}
\end{eqnarray}
while by eq.(\ref{sec.1:n5}) 
one has in the variables $\{ \overline{K}_0, j_{a\mu}, \phi_a \}$
\begin{eqnarray}
\frac{\delta \G^{(1)}}{\delta \phi_a(x)} = 0 \, .
\label{EX.2l.3.1}
\end{eqnarray}
Therefore
\begin{eqnarray}
\Delta^{(2)} & = & - \int d^Dx \,
\frac{1}{2} \frac{v_D^2}{\phi_0}
 \Theta^{-1}_{ab} \overline{\omega}_b 
\frac{\delta \G^{(1)}}{\delta \overline{K}_0(x)}  \nonumber \\
& & 
\int d^Dz \, \Big ( 
\frac{\delta \overline{K}_0(z)}{\delta \phi_a(x)} 
\frac{\delta}{\delta \overline{K}_0(z)}
+
\frac{\delta j_{c\mu}(z)}{\delta \phi_a(x)} 
\frac{\delta}{\delta j_{c\mu}(z)} 
 \Big ) 
\G^{(1)} 
\label{EX.2l.3.2}
\end{eqnarray}
It is useful to introduce two transition functions
(encoding the effect of the change of variables
from $\{ K_0, J_{a\mu}, \phi_a \}$
to $\{ \overline{K}_0, j_{a\mu}, \phi_a \}$):
\begin{eqnarray}
&& G_b(x,z) = \frac{1}{2} \frac{v_D^2}{\phi_0(x)}
 \Theta^{-1}_{ab}(x)
  \frac{\delta \overline{K}_0(z)}{\delta \phi_a(x)} \, , 
  \nonumber \\
&& H_{bc,\mu}(x,z) = \frac{1}{2} \frac{v_D^2}{\phi_0(x)}
 \Theta^{-1}_{ab}(x)
 \frac{\delta j_{c\mu}(z)}{\delta \phi_a(x)} 
\label{EX.2l.4}
\end{eqnarray}
so that eq.(\ref{EX.2l.3}) reads
\begin{eqnarray}
\Delta^{(2)} & = & - \int d^Dx \int d^Dz \, 
\overline{\omega}_b(x) \frac{\delta \G^{(1)}}{\delta \overline{K}_0(x)}  \nonumber \\
&& ~~~~ \Big ( G_b(x,z) \frac{\delta}{\delta \overline{K}_0(z)}
+ H_{bc,\mu}(x,z) \frac{\delta}{\delta j_{c\mu}(z)}  \Big )
\G^{(1)} \, .
\label{EX.2l.5}
\end{eqnarray}
In the two-loop approximation eq.(\ref{new.ho.3}) is finally 
\begin{eqnarray}
&& \!\!\!\!\!\!\!\!\!\!\!\!\!\!
\frac{\delta \G^{(2)}}{\delta \phi_b(x)} = 
\frac{\delta \Delta^{(2)}}{\delta \overline{\omega}_b(x)} 
\nonumber \\
&& \!\!\!\!\!\!\!\!
  = - \int d^Dz \, 
 \frac{\delta \G^{(1)}}{\delta \overline{K}_0(x)}  %\nonumber \\
%&& ~~~~ 
\Big ( G_b(x,z) \frac{\delta}{\delta \overline{K}_0(z)}
+ H_{bc,\mu}(x,z) \frac{\delta}{\delta j_{c\mu}(z)}  \Big )
\G^{(1)} 
\label{EX.2l.6}
\end{eqnarray}
while eq.(\ref{new.ho.4}) consequently reads
\begin{eqnarray}
&& \!\!\!\!\!\!\!\!\!\!\!\!\!\!\!\!\!\!\!\!\!\!\!\!\!
\G^{(2)}[\phi_a, K_0, J_{a\mu} ] = 
\left .
{\cal A}^{(2)}[K_0,
J_{a\mu}] \right |_{K_0 \rightarrow
\frac{1}{v_D} \overline{K}_0, J_{a\mu} \rightarrow -j_{a \mu}}
\nonumber \\
%{\cal A}^{(n)}[\frac{1}{v_D} \overline{K}_0, -j_{a\mu}]
&& \!\!\!\!\!\!\!\!\!\!\!\!\!\!\!\!\!\!\!\!\!\!\!\!\!\!\!\!\!\!\!
- \int d^Dx \, \int_0^1 dt \, \phi_b(x) \lambda_t
\int d^Dz \, 
 \frac{\delta \G^{(1)}}{\delta \overline{K}_0(x)}  %\nonumber \\
%&& ~~~~ 
\Big ( G_b(x,z) \frac{\delta}{\delta \overline{K}_0(z)}
+ H_{bc,\mu}(x,z) \frac{\delta}{\delta j_{c\mu}(z)}  \Big )
\G^{(1)} \, . 
\nonumber \\
\label{EX.new.ho.4}
\end{eqnarray}
The second line encodes the effects of the nonlinearity
of the local functional equation at two loop order. 

It should be noticed that, due to the peculiar structure of the dependence
of the one-loop vertex functional on the pions
given by eq.(\ref{sec.1:n5}),
one finds some special simplifications at two loop level.
In particular the second line of eq.(\ref{EX.new.ho.4})
does not contribute to the four point pion
Green function in the two loop approximation.

In order to show this property we remark that the expansion
of $\overline{K}_0$ starts with two $\phi$'s
while $j_{a\mu}$ starts with one $\phi$.
Hence the term with two derivatives w.r.t. $\overline{K}_0$
in the second line of eq.(\ref{EX.new.ho.4}) gives
contributions of order
$O(\phi^5)$. 

In order to obtain
the contribution to the four point pion
function of the term involving
one derivative w.r.t. $j_{c\mu}$ in the second line 
it is sufficient
to keep $H_{bc ,\mu}$ at order zero:
\begin{eqnarray}
 H_{bc;\mu}(x,z) = 
\frac{2}{v_D} \delta_{bc} \partial_{z_\mu}  \delta^D(x-z) 
+ O(\phi) \, .
\label{EX.new.1.1}
\end{eqnarray}
This yields
\begin{eqnarray}
- \frac{2}{v_D} \int d^Dx \, \phi_b(x)
\Big [ \frac{\delta \G^{(1)}}{\delta \overline{K}_0(x)} \Big ]_{\phi\phi} \partial^\mu \Big [ \frac{\delta \G^{(1)}}{\delta j_{b\mu}(x)} \Big]_{\phi} \, 
\label{EX.new.2}
\end{eqnarray}
where the subscript denotes the order of the projection
for the $\phi$'s.
Moreover the derivative
\begin{eqnarray}
\Big [ \frac{\delta \G^{(1)} }{\delta j_{b\mu}(x)} \Big ]_{\phi}
\label{EX.new.4}
\end{eqnarray}
receives contributions only from the amplitude 
$\G^{(1)}_{J_{a\mu} J_{b\nu}}$
\cite{Ferrari:2005ii}
 through
\begin{eqnarray}
&& 
\!\!\!\!\!\!\!\!\!\!\!
\frac{1}{2}\int d^Dx d^Dy  \, \G^{(1)}_{J_{a\mu}(x) J_{b\nu}(y)} 
j_{a\mu}(x) j_{b\nu}(y) \nonumber \\
&& \!\!\!\!\!\!\! = 
- \frac{1}{2} \int d^Dx d^Dy \, 
(\square g^{\mu \nu} - \partial^\mu \partial^\nu) j_{a\mu}(x)
j_{b\nu}(y) \nonumber \\
&& ~~~~~~
\int d^D p \, \frac{4i}{m_D^4} \frac{1}{D-1}
e^{i p (x-y)} I_2(p) 
\label{EX.new.5}
\end{eqnarray}
where
\begin{eqnarray}
I_2(p) = \int \frac{d^Dp}{(2\pi)^D} \, 
\frac{1}{k^2 (k+p)^2} \, .
\label{EX.new.6}
\end{eqnarray}
By taking the gradient according to eq.(\ref{EX.new.2}) one finds zero as a consequence of the transversality of
$\G^{(1)}_{J_{a\mu}(x) J_{b\nu}(y)}$. 
Therefore the second line of eq.(\ref{EX.new.ho.4})
does not give any contribution to the four point pion
function at two loop level. 
The contribution from the first line can
be derived according to the methods
discussed in Sect.~\ref{sec:oneloop}. So we get finally
\begin{eqnarray}
&& 
\!\!\!\!\!\!\!\!\!\!\!\!\!\!\!\!\!\!\!\!\!\!\!\!\!\!\!\!\!\!\!\!\!\!\!\!\!
\G^{(2)}[\phi\phi\phi\phi] = \frac{2}{v_D^4} \int %d^Dx d^Dy 
\, \G^{(2)}_{J_{a\mu}(x) J_{b\nu}(y)} 
\Big (  \partial_\mu \phi_a(x) (  - \phi_c^2(y) \partial_\nu \phi_b(y) + 2 \phi_c(y) \partial_\nu \phi_c(y) \phi_b(y)  )
\nonumber \\
&& ~~~~~~~~~~~~~~~~~~~~~~~ + \epsilon_{apq} \epsilon_{brs} \partial_\mu \phi_p(x) \phi_q(x) \partial_\nu \phi_r(y) \phi_s(y) \Big ) \nonumber \\
&& \!\!\!\!\!\! 
+ \frac{4}{v_D^4} \int %d^Dx d^Dy d^Dz 
\,  \G^{(2)}_{J_{a\mu}(x) J_{b\nu}(y) J_{c\rho}(z)}
\epsilon_{apq} \partial_\mu \phi_p(x) \phi_q(x) \partial_\nu \phi_b(y) \partial_\rho \phi_c(z) \nonumber \\
&& \!\!\!\!\!\!
+ \frac{2}{3 v_D^4}  \int %d^Dx d^Dy d^Dz d^Dw 
\,  \G^{(2)}_{J_{a\mu}(x) J_{b\nu}(y) J_{c\rho}(z) J_{d\sigma}(w)}
\partial_\mu \phi_a(x) \partial_\nu \phi_b(y) \partial_\rho \phi_c(z) \partial_\sigma \phi_d(w) \nonumber \\
&& \!\!\!\!\!\!
+ \frac{2}{v_D^3} \int \,  \G^{(2)}_{J_{a\mu}(x) J_{b\nu}(y) K_0(z)}\partial_\mu \phi_a(x) \partial_\nu \phi_b(y)
(\phi_c \square \phi_c)(z) \nonumber \\
&& \!\!\!\!\!\!
+ \frac{1}{2v_D^2} \int \, \G^{(2)}_{K_0(x) K_0(y)} (\phi_a \square \phi_a)(x) (\phi_b \square \phi_b)(y)
\, .
\label{ex.4pphi.2L}
\end{eqnarray}
%
%\spade
This formula exhibits a functional dependence of $\G^{(2)}_{\phi_{a_1} \phi_{a_2} \phi_{a_3} \phi_{a_4}}$ on the ancestor amplitudes as in the one loop
approximation (see eq.(\ref{ex.4pphi})).
This is a rather surprising result which holds as a consequence
of the transversality of the one-loop $JJ$ ancestor amplitude.
%\spade
  
\section{Hierarchy and Finite Renormalizations}\label{sec:fr}

From the results of Sects.~\ref{sec.2} and \ref{sec.3} it is clear that for any solution of the local
functional equation (\ref{sec.1:4}) the knowledge of the ancestor amplitudes order by order
in the loop expansion completely determines the dependence on the pion fields.
One important consequence of this result is that it has been obtained
without relying on the specific subtraction procedure. In particular
if we want to perform any subtraction in order to define the theory
in $D=4$, it is sufficient to operate on the ancestor amplitudes.
The subtractions on the amplitudes involving any number
of pions are induced by the integration of the functional equation
which has been developed in the previous Sections.

In this Section we exploit this property in order to shed light on the finite
renormalizations allowed 
from a mathematical point of view 
by the local symmetry and the weak power-counting theorem.

For that purpose we remark that a sufficient condition for the fulfillment of the local
functional equation (\ref{sec.1:4}) is conjectured to be (in the presence
of a symmetric regularization like Dimensional Regularization \cite{Ferrari:2005fc}) 
the validity of the same functional equation (\ref{sec.1:4}) for the functional
\begin{eqnarray}
\widehat \G = \G^{(0)} + \sum_{k=1}^\infty \widehat \G^{(k)} 
\label{fr.1}
\end{eqnarray}
where $\G^{(0)}$ is the classical action in eq.(\ref{sec.1:2}) (giving rise to the tree-level Feynman rules)
while $\widehat \G^{(k)}$ collects the $k$-th order counterterms. 
From the mathematical point of view
the latter may contain
$k$-th order finite renormalizations compatible with the symmetry properties and the weak power-counting bounds
\cite{Ferrari:2005va}.
This conjecture is supported by formal arguments \cite{paper_prep} and by some
explicit two-loop examples \cite{Ferrari:2005fc}.

We will now prove that the ancestor amplitudes of $\widehat \G$ can be obtained from the tree-level
ancestor amplitudes through a suitable redefinition of the classical sources $J_{a\mu}$ and $K_0$:
\begin{eqnarray}
&& J_{a\mu} \rightarrow J_{a\mu} + A_{1,a\mu}(J) + A_{2,a\mu}(J) + \dots \, , \nonumber \\
&& K_0 \rightarrow K_0(1 + B_1(K_0,J) + B_2(K_0,J) + \dots) 
\label{fr.2}
\end{eqnarray}
where $A_{j,a\mu}, B_j$ are of order $\hbar^j$. $A_{j,a\mu}$ does not depend on $K_0$.
We also set $A_{0,a\mu} = J_{a\mu}, B_0=1$.

First we notice that by using integration by parts it is always possible to decompose in a unique way
an integrated local functional $\int d^Dx \, X(J,K_0)$ according to
\begin{eqnarray}
\int d^Dx \, X(J,K_0) = \int d^Dx \, \Big ( J_{a\mu} {\cal P}_a^\mu[X] + K_0 {\cal Q}[X] \Big ) 
\label{fr.3}
\end{eqnarray}
where ${\cal P}_a^\mu[X]$ 
is the result of the projection of $X$ into a local
function of $J$ and its derivatives while ${\cal Q}[X]$ 
includes also local dependence on $K_0$ and its derivatives.
%Notice that 
%%
%\begin{eqnarray}
%{\cal P}_a^\mu (J_{b\nu} {\cal P}_b^\nu (X) ) = {\cal P}_a^\mu (X) \, .
%\label{fr.4}
%\end{eqnarray}
%
In order to determine the unknown functions $A_{j,a\mu}$ and $B_j$ in eq.(\ref{fr.2}) 
we perform the substitution (\ref{fr.2}) into 
\begin{eqnarray}
&& \G^{(0)}[0,J_{a\mu},K_0] = \int d^Dx \, \Big (  \frac{v_D^2}{8} J^2 + v_D K_0 \Big ) \nonumber \\
&& ~~~~~~~~~ \rightarrow \sum_{l=0}^\infty 
\int d^Dx \, \Big (  \frac{v_D^2}{8} \sum_{j=0}^l A_{j,a\mu} A_{l-j,a}^\mu   + v_D K_0 B_l \Big )
\label{fr.5}
\end{eqnarray}
and then compare the second line of the above equation with the ancestor 
counterterms
$$\widehat \G^{(l)}[0,K_0,J_{a\mu}] \equiv \int d^Dx \ \widehat{\cal L}_l(J,K_0) \, .$$
This gives
\begin{eqnarray}
&& 
\!\!\!\!\!\!\!\!\!\!\!\!\!\!\!\!\!\!\!\!\!\!\!\!\!\!\!\!\!\!\!
\int d^Dx \, \widehat {\cal L}_l = \frac{v_D^2}{8} \int d^Dx  \sum_{j=0}^l A_{j,a\mu} A_{l-j,a}^\mu  
+ \int d^Dx \, v_D K_0  B_l 
\nonumber \\
&&
\!\!\!\!\!\!\!\!\!\!\!\!\!\!\!\!\!\!\!\!\!\!\!\!\!
 = \int d^Dx \, \Big ( \frac{v_D^2}{4} J_{a\mu} A_{l,a}^\mu + \frac{v_D^2}{8} \sum_{j=1}^{l-1} A_{j,a\mu} A_{l-j,a}^\mu + v_D K_0  B_l \Big ) \, , 
              ~~ l=1,2,3,\dots
\label{fr.6}
\end{eqnarray}
and hence we derive the recursive solution
\begin{eqnarray}
&&
\!\!\!\!\!\!\!\!\!\!\!\!
B_0 = 1 \, , ~~~~ B_l = \frac{1}{v_D} {\cal Q} [\widehat {\cal L}_l] \,  , \nonumber \\
&& 
\!\!\!\!\!\!\!\!\!\!\!\!
A_{0,a\mu} = J_{a\mu} \, , \nonumber \\
&& 
\!\!\!\!\!\!\!\!\!\!\!\!
A_{l,a\mu} = \frac{4}{v_D^2} {\cal P}_{a\mu} [\widehat {\cal L}_l] - \frac{1}{2} {\cal P}_{a\mu}
\Big [ \sum_{j=1}^{l-1} A_{j,b\nu} A_{l-j,b}^\nu \Big ]  \, , ~~~~ l=1,2,3,\dots        
\label{fr.7}
\end{eqnarray}
This result  states that all possible finite renormalizations
in $\widehat \G^{(k)}$, $k>1$, compatible with the local symmetry and the weak power-counting, 
can in fact be interpreted as a redefinition of the sources $J_{a\mu}$ and $K_0$ by finite quantum corrections.
The latter correspond to the ambiguities allowed
in the effective field theory approach discussed
in \cite{paper_prep}.

\section{Conclusions}\label{sec.5}

The requirement of the invariance of the group
Haar measure under local left multiplication
can be implemented by a local functional equation
for the 1-PI vertex functional of the nonlinear
sigma model.
This equation can be preserved by the subtraction
procedure and completely fixes the dependence of
the vertex functional on the pion fields
in terms of the ancestor amplitudes (i.e.
amplitudes only involving the flat connection
and the nonlinear sigma model constraint).

Very remarkably the recursive solution
can be written in a very compact form 
in terms of invariant variables 
(inducing an implicit dependence of 
the vertex functional on the quantized field)
plus (at order $n \geq 2$)
a contribution yielding an explicit dependence
on $\phi_a$. The latter is
fixed by lower order terms (see eq.(\ref{new.ho.4}))
and does not affect the $n$-th loop ancestor amplitudes.
This solution provides the full dependence 
of the 1-PI symmetric amplitudes on the pion fields.

From a technical point of view the method which
has been developed in order to integrate
the local functional equation extends 
the cohomological techniques originally developed
in the context of gauge theories. In particular 
it deals with the full Green functions of the theory
(no locality restrictions) and it solves explicitly
the inhomogeneous equation (arising from the loop
expansion of the bilinear local functional equation)
in the absence of multiplicative renormalization (as
it happens for the subtraction procedure of 
the nonlinear sigma model).

The integration of the local functional equation
at higher orders in the loop expansion
allows to treat a new class of problems which
could not be addressed by the knowledge
of the solutions of the linearized functional equation only.

Among them we think that two issues are worthwhile to be
pointed out. 
The first one is that our method
allows the
determination of all pion amplitudes  at higher orders
in Chiral Perturbation Theory.

The second one is the possibility
to investigate the use of the techniques discussed in this paper in order 
to set up a consistent framework for the study of the structure of the higher order
divergences within the program of the quantization
of the St\"uckelberg model for non-abelian massive
gauge bosons.

\section*{Acknowledgments}

One of us (A.Q.) would like to thank G.~Barnich and
G.~Colangelo for useful discussions. He also
acknowledges the warm hospitality at the
Institut f\"ur Theoretische Physik in Bern, Switzerland.
\appendix
\section{Consistency condition}\label{appA}

In this appendix we verify eq.(\ref{ho.cc}) as a consequence of the recursive validity
of the functional equation at lower orders.
The technique is a variant of the general proof of the consistency
condition in the Batalin-Vilkovisky (BV) formalism \cite{Gomis:1994he}. One should notice
that in the present case the introduction of the antifield $J_{a\mu}^*$ for the background source
$J_{a\mu}$ is forbidden (since this would lead to an empty cohomology \cite{Henneaux:1998hq}). Therefore
one cannot use the standard BV bracket.

\medskip
The local functional equation at order $n$ in the loop expansion reads
\begin{eqnarray}
s \G^{(n)} = - \frac{1}{2} \sum_{j=1}^{n-1} \int d^Dx \, \omega_a \frac{\delta \G^{(j)}}{\delta K_0}
                                                         \frac{\delta \G^{(n-j)}}{\delta \phi_a} \, , 
\label{wz.1}
\end{eqnarray}
which is useful to rewrite in the more symmetric form
\begin{eqnarray}
s \G^{(n)} & = &  
%- \frac{1}{2}\sum_{j=1}^{n-1} \Big (  (\G^{(j)}, \G^{(n-j)}) 
%                                    + (\G^{(n-j)}, \G^{(j)}) \Big ) 
%\nonumber \\
%& = & 
- \frac{1}{2} \sum_{j=1}^{n-1} \langle \G^{(j)}, \G^{(n-j)} \rangle 
\, .
\label{wz.2}
\end{eqnarray}
In the above equation we have adopted the notation
\begin{eqnarray}
\langle X, Y \rangle & = & %(X,Y) + (Y,X) \nonumber \\ 
                     %& = &  
                            \int d^Dx \, \frac{1}{2} \omega_a \Big (
                                         \frac{\delta X}{\delta K_0}
                                         \frac{\delta Y}{\delta \phi_a}
                                       + \frac{\delta Y}{\delta K_0}
                                         \frac{\delta X}{\delta \phi_a} 
					 \Big ) \, .
\label{wz.3}
\end{eqnarray}
The following properties hold for $\langle X, Y \rangle$:
\begin{eqnarray}
&& \langle X, Y \rangle = \langle Y, X \rangle \, , \nonumber \\
&& s \langle X, Y \rangle = - \langle s X, Y \rangle
- \langle X, s Y \rangle ~~~~~ X,Y ~~~ \mbox{bosonic} \, .
\label{wz.4}
\end{eqnarray}

\medskip
We denote by $\Delta^{(n)}$ the R.H.S. of eq.(\ref{wz.2}), i.e. we
set
\begin{eqnarray}
\Delta^{(n)} =  - \frac{1}{2} \sum_{j=1}^{n-1} \langle \G^{(j)}, \G^{(n-j)} \rangle \, .
\label{wz.5}
\end{eqnarray}
If a solution to eq.(\ref{wz.2}) exists, by the nilpotency of $s$ 
the following consistency condition has to be verified:
\begin{eqnarray} 
s \Delta^{(n)} = 0 \, .
\label{wz.n.1}
\end{eqnarray}
Let us verify that this is indeed the case under the recursive
assumption that the master equation has been fulfilled
up to order $n-1$.

By using eq.(\ref{wz.4}) we get
\begin{eqnarray}
s \Delta^{(n)} & = & s \Big ( -\frac{1}{2} \sum_{j=1}^{n-1}
\langle \G^{(j)}, \G^{(n-j)} \rangle \Big ) \nonumber \\
             & = & + \frac{1}{2}  \sum_{j=1}^{n-1}
                   \Big ( \langle s \G^{(j)}, \G^{(n-j)} \rangle
                        + \langle \G^{(j)}, s \G^{(n-j)} \rangle \Big )
\nonumber \\
             & = &  + \frac{1}{2}  \sum_{j=1}^{n-1}
                   \Big ( \langle s \G^{(j)}, \G^{(n-j)} \rangle
                        + \langle  s \G^{(n-j)},  \G^{(j)} \rangle \Big )
\nonumber \\
             & = & \sum_{j=1}^{n-1} \langle s \G^{(j)}, \G^{(n-j)} \rangle
\label{wz.6}
\end{eqnarray}
Now we use the recursive assumption that
\begin{eqnarray}
s \G^{(j)} & = & - \frac{1}{2} \sum_{k=1}^{j-1} \langle \G^{(k)}, \G^{(j-k)} \rangle 
\label{wz.7}
\end{eqnarray}
so that
\begin{eqnarray}
s \Delta^{(n)} & = & -\frac{1}{2} \sum_{j=1}^{n-1} \sum_{k=1}^{j-1}
                       \langle \langle \G^{(k)}, \G^{(j-k)} \rangle , \G^{(n-j)} \rangle \nonumber \\
                & = & -\frac{1}{2} \cdot \frac{1}{3} \sum_{j=1}^{n-1} \sum_{k=1}^{j-1} \Big ( \langle \langle \G^{(k)}, \G^{(j-k)} \rangle , \G^{(n-j)} \rangle
       +\langle \langle \G^{(j-k)}, \G^{(n-j)} \rangle , \G^{(k)} \rangle
\nonumber \\
&& ~~~~~~~~~~~~~~~~~~~~~~ 
       +\langle \langle \G^{(n-j)}, \G^{(k)} \rangle , \G^{(j-k)} \rangle
 \Big ) \, .
\label{wz.8}
\end{eqnarray}
It turns out that the symmetrized bracket enjoys the following Jacobi identity
($X,Y,Z$ are assumed to be bosonic):
\begin{eqnarray}
\langle \langle X,Y \rangle, Z \rangle + 
\langle \langle Z,X \rangle, Y \rangle +
\langle \langle Y,Z \rangle, X \rangle = 0 \, .
\label{wz.9}
\end{eqnarray}
The proof of the above equation is provided in the next subsection.
By using eq.(\ref{wz.9}) into eq.(\ref{wz.8}) we  finally get
\begin{eqnarray}
s \Delta^{(n)} = 0 \, .
\label{wz.10}
\end{eqnarray}

\subsection{Proof of the Jacobi identity for the symmetrized bracket}

We assume $X,Y,Z$ to be bosonic.
We write explicitly $\langle \langle X,Y \rangle , Z \rangle$:
\begin{eqnarray}
\langle \langle X,Y \rangle , Z \rangle & = & 
       \int d^Dx \, \frac{1}{2} \omega_a(x) \frac{\delta}{\delta K_0(x)}
       ( \langle X,Y \rangle ) \frac{\delta Z}{\delta \phi_a(x)}
     \nonumber \\
     & &  
     + \int d^Dx \, \frac{1}{2} \omega_a(x) \frac{\delta Z}{\delta K_0(x)}
       \frac{\delta}{\delta \phi_a(x)} (\langle X,Y \rangle)
     \nonumber \\
     & = & \int d^Dx \frac{1}{2} \omega_a(x) \frac{\delta}{\delta K_0(x)}
           \Big [ \int d^Dy \, \frac{1}{2} \omega_b(y) \frac{\delta X}{\delta K_0(y)}  
                                                    \frac{\delta Y}{\delta \phi_b(y)} \nonumber \\
     &   & ~~~~~ +\int d^Dy \, \frac{1}{2} \omega_b(y) \frac{\delta Y}{\delta K_0(y)} 
                  \frac{\delta X}{\delta \phi_b(y)} \Big ] \frac{\delta Z}{\delta \phi_a(x)} \nonumber \\
     &   & + \int d^Dx \, \frac{1}{2} \omega_a(x) \frac{\delta Z}{\delta K_0(x)}
                          \frac{\delta}{\delta \phi_a(x)} 
                          \Big [ \int d^Dy \, \frac{1}{2} \omega_b(y) 
                                 \frac{\delta X}{\delta K_0(y)}\frac{\delta Y}{\delta \phi_b(y)} \nonumber \\
     &   & ~~~~~ + \int d^Dy \, \frac{1}{2} \omega_b(y) 
                                 \frac{\delta Y}{\delta K_0(y)}\frac{\delta X}{\delta \phi_b(y)} \Big ]
\label{wz.11}
\end{eqnarray}
We notice that the following terms in the R.H.S. of eq.(\ref{wz.11})
\begin{eqnarray}
&& \int d^Dx d^Dy \, \frac{1}{2} \omega_a(x) \frac{1}{2} \omega_b(y) 
                     \frac{\delta Z}{\delta K_0(x)} \frac{\delta X}{\delta K_0(y)} \frac{\delta^2 Y}{\delta \phi_a(x) \delta \phi_b(y)} \, , \nonumber \\
&& \int d^Dx d^Dy \, \frac{1}{2} \omega_a(x) \frac{1}{2} \omega_b(y) 
                     \frac{\delta Z}{\delta K_0(x)} \frac{\delta Y}{\delta K_0(y)} \frac{\delta^2 X}{\delta \phi_a(x) \delta \phi_b(y)} 
\label{wz.12}
\end{eqnarray}
are zero since $\omega_a(x)$ and $\omega_b(y)$ are anticommuting.

\medskip
We make use of eq.(\ref{wz.11}) in order to write the sum
$ \langle \langle X, Y \rangle, Z \rangle + \mbox{cyclic}$. We organize the terms
according to the number of derivatives w.r.t $K_0$ acting on a single functional.
We obtain
\begin{eqnarray}
\langle \langle X, Y \rangle, Z \rangle + \mbox{cyclic} & = & \int d^Dx \int d^Dy ~
\frac{1}{2} \omega_a(x) \frac{1}{2} \omega_b(y) \nonumber \\
&& ~~~~~ \times \Big [
\frac{\delta^2 X}{\delta K_0(x) \delta K_0(y)} 
\Big ( \frac{\delta Y}{\delta \phi_b(y)} \frac{\delta Z}{\delta \phi_a(x)} 
      +\frac{\delta Z}{\delta \phi_b(y)} \frac{\delta Y}{\delta \phi_a(x)} \Big ) \nonumber \\
&& ~~~~~~~ + 
\frac{\delta^2 Y}{\delta K_0(x) \delta K_0(y)} 
\Big ( \frac{\delta Z}{\delta \phi_b(y)} \frac{\delta X}{\delta \phi_a(x)} 
      +\frac{\delta X}{\delta \phi_b(y)} \frac{\delta Z}{\delta \phi_a(x)} \Big ) \nonumber \\
&& ~~~~~~~ + 
\frac{\delta^2 Z}{\delta K_0(x) \delta K_0(y)} 
\Big ( \frac{\delta X}{\delta \phi_b(y)} \frac{\delta Y}{\delta \phi_a(x)} 
      +\frac{\delta Y}{\delta \phi_b(y)} \frac{\delta X}{\delta \phi_a(x)} \Big ) \Big ] \nonumber \\
&& + \int d^Dx \int d^Dy ~
\frac{1}{2} \omega_a(x) \frac{1}{2} \omega_b(y) \nonumber \\
&& ~~~~~ \times \Big [ \frac{\delta X}{\delta K_0(y)} \Big ( \frac{\delta^2 Y}{\delta K_0(x) \delta \phi_b(y)}
\frac{\delta Z}{\delta \phi_a(x)} + \frac{\delta^2 Z}{\delta K_0(x) \delta \phi_b(y)} \frac{\delta Y}{\delta \phi_a(x)} \Big ) \nonumber \\
&& ~~~~~~~~~ + \frac{\delta X}{\delta K_0(x)} \Big ( \frac{\delta^2 Y}{\delta \phi_a(x) \delta K_0(y)}
\frac{\delta Z}{\delta \phi_b(y)} + \frac{\delta^2 Z}{\delta \phi_a(x) \delta K_0(y)}
\frac{\delta Y}{\delta \phi_b(y)} \Big ) \nonumber \\
&& ~~~~~~~~~  + \mbox{cyclic} \Big ]
\label{wz.13}
\end{eqnarray}
The terms in the first block between square brackets in the above equation
vanish by symmetry once the anticommutativity of $\omega_a(x), \omega_b(y)$ is taken
into account. 

The second block requires some manipulations. If one exchanges $y \leftrightarrow x$
and $a \leftrightarrow b$ in the second line of the second block, the latter becomes
\begin{eqnarray}
&& + \int d^Dx \int d^Dy ~
\frac{1}{2} \omega_a(x) \frac{1}{2} \omega_b(y) \nonumber \\
&& ~~~~~ \times  \frac{\delta X}{\delta K_0(y)} \Big ( \frac{\delta^2 Y}{\delta K_0(x) \delta \phi_b(y)}
\frac{\delta Z}{\delta \phi_a(x)} + \frac{\delta^2 Z}{\delta K_0(x) \delta \phi_b(y)} \frac{\delta Y}{\delta \phi_a(x)} \Big ) \nonumber \\
&& + \int d^Dx \int d^Dy ~
\frac{1}{2} \omega_b(y) \frac{1}{2} \omega_a(x) \nonumber \\
&& ~~~~~ \times  \frac{\delta X}{\delta K_0(y)} \Big ( \frac{\delta^2 Y}{\delta K_0(x) \delta \phi_b(y)}
\frac{\delta Z}{\delta \phi_a(x)} + \frac{\delta^2 Z}{\delta K_0(x) \delta \phi_b(y)} \frac{\delta Y}{\delta \phi_a(x)} \Big ) \nonumber \\
&& ~~~~~~~~~  + \mbox{cyclic} 
\label{wz.14}
\end{eqnarray}
The above expression is zero since $\omega_a(x), \omega_b(y)$ anticommute.

\medskip
Therefore we establish the Jacobi identity for the symmetrized bracket in the form
\begin{eqnarray}
\langle \langle X,Y \rangle, Z \rangle + 
\langle \langle Z,X \rangle, Y \rangle +
\langle \langle Y,Z \rangle, X \rangle = 0 
\label{wz.15}
\end{eqnarray}
with $X,Y,Z$ bosonic.

\section{Integrability condition}\label{app:int}

In this Appendix we check that eq.(\ref{new.ho.3.2}) holds as a consequence of eq.(\ref{ho.cc}).
Eq.(\ref{ho.cc}) reads in the variables $\{ \overline{K}_0, j_{a\mu}, \phi_a, \overline{\omega}_a\}$ 
\begin{eqnarray}
\int d^Dx \, \overline{\omega}_a(x) \frac{\delta \Delta^{(n)}}{\delta \phi_a(x)} = 0 \, .
\label{app:int.1}
\end{eqnarray}
By differentiating the above equation w.r.t. $\overline{\omega}_a(x), \overline{\omega}_b(y)$ and by setting
$\overline{\omega}=0$ we get
\begin{eqnarray}
\frac{\delta^2 \Delta^{(n)}}{\delta \overline \omega_b(y) \delta \phi_a(x)} 
= \frac{\delta^2 \Delta^{(n)}}{\delta \overline \omega_a(x) \delta \phi_b(y)} \, .
\label{app:int.2}
\end{eqnarray}
Let us now consider the R.H.S. of eq.(\ref{new.ho.3.2}). For each permutation $\sigma \in S_m$ 
there exists a unique integer $1 \leq K \leq m$ such that $\sigma(K)=1$.
Therefore (we drop here the dependence on $\zeta_1, \dots, \zeta_n$ since the latter does not play any role
in the following argument)
\begin{eqnarray}
&& 
\!\!\!\!\!\!\!\!\!\!\!\!\!\!\!\!\!\!\!\!\!\!\!\!
\frac{1}{m!} \sum_{\sigma \in S_m} \Delta^{(n)}_{\overline{\omega}_{a_{\sigma(1)}} \phi_{a_{\sigma(2)}} \dots \phi_{a_{\sigma(m)}}}
 =  \frac{1}{m!} \sum_{\sigma \in S_{m-1}[{2,\dots, m}]} \Delta^{(n)}_{\overline{\omega}_{a_1} \phi_{a_{\sigma(2)}} \dots \phi_{a_{\sigma(m)}}}
\nonumber \\
&& 
+ \frac{1}{m!} \sum_{K=2}^m  \sum_{\sigma \in S_{m-1}[{1,2,\dots,\widehat{K}, \dots, m}]} \Delta^{(n)}_{\overline{\omega}_{a_{\sigma(1)}}
\phi_{a_{\sigma(K)}} \phi_{a_{\sigma(2)}} \dots  \widehat{\phi}_{a_{\sigma(K)}} \dots \phi_{a_{\sigma(m)}}} \, .
\label{app:int.3}
\end{eqnarray}
In the above equation a hat over a variable denotes omission of the latter from the relevant list and $S_{m-1}[a,b,\dots,c]$
denotes the group of permutations over the $m-1$ elements $\{ a,b,\dots,c \}$.

We now use eq.(\ref{app:int.2}) in the second line of eq.(\ref{app:int.3}) as well as the fact that $\sigma(K)=1$ and we get
\begin{eqnarray}
&& 
\!\!\!\!\!\!\!\!\!\!\!\!\!\!\!\!\!\!\!\!\!\!\!\!
\frac{1}{m!} \sum_{\sigma \in S_m} \Delta^{(n)}_{\overline{\omega}_{a_{\sigma(1)}} \phi_{a_{\sigma(2)}} \dots \phi_{a_{\sigma(m)}}}
 =  \frac{1}{m!} \sum_{\sigma \in S_{m-1}[{2,\dots, m}]} \Delta^{(n)}_{\overline{\omega}_{a_1} \phi_{a_{\sigma(2)}} \dots \phi_{a_{\sigma(m)}}}
\nonumber \\
&& 
+ \frac{1}{m!} \sum_{K=2}^m  \sum_{\sigma \in S_{m-1}[{1,2,\dots,\widehat{K}, \dots, m}]} \Delta^{(n)}_{\overline{\omega}_{a_1}
\phi_{a_{\sigma(1)}} \phi_{a_{\sigma(2)}} \dots  \widehat{\phi}_{a_{\sigma(K)}} \dots \phi_{a_{\sigma(m)}}} \, .
\label{app:int.4}
\end{eqnarray}
By the Bose statistics of the $\phi$'s we also get
\begin{eqnarray}
&& 
\!\!\!\!\!\!\!\!\!\!\!\!\!\!\!\!\!\!\!\!\!\!\!\!\!\!\!\!\!\!\!\!
\frac{1}{m!} \sum_{\sigma \in S_m} \Delta^{(n)}_{\overline{\omega}_{a_{\sigma(1)}} \phi_{a_{\sigma(2)}} \dots \phi_{a_{\sigma(m)}}}
 =  \frac{1}{m}  \Delta^{(n)}_{\overline{\omega}_{a_1} \phi_{a_2} \dots \phi_{a_m}}
%\nonumber \\
%&& 
+ \frac{m-1}{m} \Delta^{(n)}_{\overline{\omega}_{a_1}
\phi_{a_2} \phi_{a_3} \dots  \dots \phi_{a_m}} \nonumber \\
&& = \Delta^{(n)}_{\overline{\omega}_{a_1} \phi_{a_2} \dots \phi_{a_m}} \, ,
\label{app:int.5}
\end{eqnarray}
which proves eq.(\ref{new.ho.3.2}).

A comment is in order here. It is a well-known fact in cohomological algebra 
\cite{Piguet:1995er, Barnich:2000zw, Quadri:2002nh}
 that if a local functional
with ghost number one
satisfies the consistency condition in eq.(\ref{ho.cc}) (i.e. it is BRST closed) and the BRST differential $s$ 
has been trivialized by reduction to a doublet pair 
$$s\phi_a = \overline{\omega}_a, ~~~~ s\overline{\omega}_a=0$$
then that functional is also BRST-exact. 

The present analysis
generalizes this result to the case of arbitrary functionals, the locality property being nowhere used in the
above construction.

\section{One-loop invariants}\label{app:B}

We report here the invariants parameterizing the one-loop divergences
of the nonlinear sigma model in $D=4$ \cite{Ferrari:2005va}.
The background connection is denoted by $J_{a\mu}$. 
\begin{eqnarray}
&& {\cal I}_1 = \int d^Dx \, \Big [ D_\mu ( F -J )_\nu \Big ]_a \Big [ D^\mu ( F -J )^\nu \Big ]_a  \, , 
\nonumber \\
&& {\cal I}_2 = \int d^Dx \, \Big [ D_\mu ( F -J )^\mu \Big ]_a \Big [ D_\nu ( F -J )^\nu \Big ]_a  \, , 
\nonumber \\
&& {\cal I}_3 = \int d^Dx \, \epsilon_{abc} \Big [ D_\mu ( F -J )_\nu \Big ]_a \Big ( F^\mu_b -J^\mu_b \Big ) \Big ( F^\nu_c -J^\nu_c \Big ) \, ,  \nonumber \\
&& {\cal I}_4 = \int d^Dx \, \Big ( \frac{m_D^2 K_0}{\phi_0} - \phi_a \frac{\delta S_0}{\delta \phi_a} \Big )^2 \, , \nonumber \\
&& {\cal I}_5 = \int d^Dx \, \Big ( \frac{m_D^2 K_0}{\phi_0} - \phi_a \frac{\delta S_0}{\delta \phi_a} \Big ) \Big ( F^\mu_b -J^\mu_b \Big )^2 \, , 
\nonumber \\
&& {\cal I}_6 = \int d^Dx \, \Big ( F^\mu_a -J^\mu_a\Big  )^2
 \Big ( F^\nu_b -J^\nu_b \Big )^2 \, , \nonumber \\
&& {\cal I}_7 = \int d^Dx \, \Big ( F^\mu_a -J^\mu_a\Big  )
   \Big ( F^\nu_a -J^\nu_a\Big  ) 
%\nonumber \\&& ~~~~~~~~~~~~~~~~
   \Big ( F_{b\mu} -J_{b\mu} \Big  )
   \Big ( F_{b\nu} -J_{b\nu} \Big  ) \, .
\label{appE:4}
\end{eqnarray}
In the above equation $D_\mu[F]$ stands for the covariant derivative w.r.t. $F_{a\mu}$
\begin{eqnarray}
D_\mu[F]_{ab}  = \delta_{ab} \partial_\mu + \epsilon_{acb} F_{c \mu} \, .
\label{appE.16}
\end{eqnarray}

\end{document}